\documentclass[final,3p,times,twocolumn]{elsarticle}

\usepackage{amssymb}
\usepackage{amsmath}
\usepackage{xcolor}
\usepackage{multirow}
\usepackage[hidelinks]{hyperref}

\journal{Journal of Computational Physics}

\begin{document}

\begin{frontmatter}



\title{Fill the gaps: continuous in time interpolation of fluid dynamical simulations}


\author[1]{Jonas Pronk} 
\author[1]{Oliver Porth}
\author[2,3]{Jordy Davelaar}

\affiliation[1]{organization={Anton Pannekoek Institute for Astonomy, University of Amsterdam},
            addressline={Science Park 904}, 
            city={Amsterdam},
            postcode={1098 XH}, 
            country={The Netherlands}}

\affiliation[2]{organization={Department of Astrophysical Sciences, Peyton Hall, Princeton University},
            city={Princeton},
            postcode={08544},
            state={New Jersey},
            country={USA}}
            
\affiliation[3]{organization={NASA Hubble Fellowship Program, Einstein Fellow}}
\begin{abstract}
Flexible and accurate interpolation schemes using machine learning could be of great benefit for many use-cases in numerical simulations and post-processing, such as temporal upsampling or storage reduction. In this work, we adapt the physics-informed token transformer (PITT) network for multi-channel data and couple it with Fourier neural operator (FNO). The resulting PITT FNO network is trained for interpolation tasks on a dataset governed by the Euler equations. We compare the performance of our machine learning model with a linear interpolation baseline and show that it requires $\sim$6-10 times less data to achieve the same mean square error of the interpolated quantities. Additionally, PITT FNO has excellent mass and energy conservation as a result of its physics-informed nature. We further discuss the ability of the network to recover fine detail using a spectral analysis.  Our results suggest that loss of fine details is related to the decreasing correlation time of the data with increasing Fourier mode which cannot be resolved by simply increasing Fourier mode truncation in FNO.
\end{abstract}




\end{frontmatter}


\section{introduction}
Simulations in fluid dynamics often involve expensive computations and large datasets that continuously push the limits of our software and hardware capabilities. Machine learning (ML) techniques are now emerging as a promising alternative to solve partial differential equations (PDEs) \cite{brunton2020machine}. In many cases, ML algorithms provide increased flexibility and computational efficiency, allowing them to handle increasingly versatile and sizeable datasets. These qualities, together with recent innovations such as foundational models, have triggered the interest of many scientists and given a significant boost to the use of ML in fluid dynamics.  \\

In fluid dynamics, focus often lies on temporal extrapolation, evolving fluids from initial conditions. Many types of neural networks (NNs) have been developed for this task. As NNs are thought of as universal function estimators, they provide a great approximation to many functions \cite{bolcskei2019optimal}. In recent years, convolutional neural network (CNN) architectures have been widely explored in fluid dynamics to learn fluid flows \cite{pathak2020using, wang2020towards, stachenfeld2021learned, shi2015convolutional}. Examples include TF-Net \cite{wang2020towards}, which aims to forecast turbulent flows across multiple scales through spectral decomposition, and ConvLSTM \cite{shi2015convolutional}, which implements a convolutional structure into a long short-term memory (LSTM) model. Physics-informed neural networks (PINNs) aim to satisfy conservation laws by integrating data and mathematical models. PINNs have been successfully applied to solve PDE systems, such as the Navier-Stokes equations \cite{jin2021nsfnets, baymani2015artificial}, and the Euler equations \cite{mao2020physics}.\\

Neural operators and transformer-based models have also been developed for extrapolation tasks.  Examples of neural operators are DeepONet \cite{lu2019deeponet}, OFormer \cite{li2022transformer}, and Fourier Neural Operator (FNO) \cite{li2020fourier}. By parametrising the integral kernel in Fourier space, FNO reduces the computational cost of evaluating integral operators. A physics-informed neural operator (PINO) model was developed, which combines the strengths of both FNO and PINNs to accelerate the training process of the neural operator \cite{li2024physics}. Physics-informed token transformer (PITT) is a transformer-based model that builds forth on existing neural operator architectures by implementing information on the underlying PDE into the neural operator output \cite{lorsung2024physics}. Furthermore, the embedding of the time difference in PITT allows for the use of arbitrary timesteps and is therefore capable of continuous-in-time predictions. Transformers have been used in fluid dynamics to process temporal information in order to predict flow dynamics \cite{geneva2022transformers, han2022predicting}, or to capture structures in the spatial domain \cite{kissas2022learning, li2022transformer}. With rare exceptions \cite{roy2024attention}, these networks are used to advance the fluid further in time while the evaluation for interpolation tasks is rarely considered so far.\\

Flexible and accurate interpolation schemes using ML could be of great benefit for many use-cases of post-processing. One such concrete use-case are raytracing calculations that involve the finite propagation of the speed of light, often referred to as 'slow-light' \cite{saiz2025probing, vos2024magnetic}. These calculations play an important role in modelling emission from compact objects, such as black holes and neutron stars \cite{bronzwaer2018raptor, abuter2018detection, das2024pulse, choudhury2024exploring}. During these calculations, high cadence is required for accurate interpolation of fluid variables between the time snapshots \cite{dexter2010submillimeter}. Implementing slow-light in raytracing therefore requires hundreds of simulation snapshots to be loaded onto memory. Depending on the resolution and output frequency, the required disk storage can range from hundreds of GB to a few TB of data. The computational challenges associated with slow-light mean that it is often ignored \cite{bronzwaer2018raptor, vos2024magnetic}. However, ignoring slow-light limits our ability to investigate causal relationships between emission components and properly address multi-wavelength source variability. It is hence appealing to learn fluid dynamics between coarsely sampled snapshots to provide an accurate and constrained interpolant.  
While slow-light raytracing is one use-case example, the method can be applied to various other use-cases in post-processing and potentially as an on-the-fly application.  

In post-processing, the methods presented here will be useful whenever the cadence required for accurate post-processing of fluid simulations exceeds the actual output frequency of simulation snapshots.
For an on-the-fly use-case, we can imagine applications where stiff operators (e.g. radiative cooling, Ampere's law in the low-resistivity regime of relativistic resistive MHD \cite{RipperdaBacchiniEtAl2019,LierJainEtAl2025}) are split from non-stiff operators (e.g. numerical flux update) which are evolved on very different timesteps.  The machine-learning model can be used to learn the dynamics of the non-stiff operator which can be evolved with low cadence and present a finely sampled solution for sub-stepping in the stiff operator.  
\\

Although models for interpolation tasks remain underexplored in fluid dynamics, they have been widely explored in the computer vision field for video interpolation. Here networks have been utilized to create slow motion effects or to facilitate smooth motion between images. These networks employ a wide variety of architectures to accurately predict an intermediate image between the input frames. Most current interpolation networks use CNN-based methods \cite{long2016learning, niklaus2017video, jiang2018super, sun2018pwc, bao2019depth, niklaus2020softmax}, although more recently transformer-based methods have seen a steady introduction \cite{shi2022video, lu2022video}. Many of these networks use optical flow learning methods to estimate the motion between the input frames, and warp the input frames to generate the intermediate frame \cite{niklaus2020softmax, jiang2018super}. To obtain visually pleasing results, some networks extract image features at multiple resolution levels, allowing them to predict motion at coarse scales as well as fine scales \cite{reda2022film, sun2018pwc}. The challenges faced in video interpolation do not always align with those in physics. Where the field of fluid dynamics is generally concerned with physicality of the model in the form of physical or conservation laws, the computer vision field is more concerned with image sharpness or artifacts caused by occlusions. Furthermore, these models can typically only be applied to three channels (RGB), and do not provide continuous-in-time capabilities.\\

In this work, we explore the use of machine learning architectures to learn the interpolation task for a challenging use-case of a 2D shock dominated flow. The features we desire for this method are the following:
\begin{enumerate}
    \item Multi-channel support, allowing the network to exploit cross-channel correlations
    \item Continuous in time predictions with arbitrary timesteps
\end{enumerate}
These model characteristics are acquired through the use of the FNO and PITT networks. These networks have shown promising results on 1D and 2D benchmarks.
FNO was originally trained and tested on a set of Navier-Stokes simulations, showing improved performance over other ML models \cite{he2016deep,ronneberger2015u,wang2020towards}. The application of FNO to magnetohydrodynamical simulation in the form of an Orszag–Tang vortex was explored, reducing errors 97\% compared to an UNet baseline while providing a 25$\times$ inference speed-up compared to a high-order finite-volume solver \cite{duarte2025spectral}. In recent years, many variants of FNO have been developed with applications in geoscience \cite{wen2022u} and weather forecasting \cite{guibas2021adaptive, pathak2022fourcastnet}. 
The use of PITT for interpolation has been explored by coupling it with a residual neural network in FluidsFormer \cite{roy2024attention}. However, this network was designed with the aim of creating visually appealing and smooth animations to be used for visual effects. We will be evaluating the accuracy and physicality of FNO and PITT for continuous-in-time interpolation. Furthermore, we will be adapting the PITT network to handle multi-channel data. This paper is structured as follows. In Section \ref{sec:methods}, we provide an overview of the network architectures, Section \ref{sec:training} discusses our sampling an training process, Section \ref{sec:results} quantifies and compares network performance. We finalise with a discussion in Section \ref{sec:discussion} and the conclusion in Section \ref{sec:conclusion}.

\section{Methods}\label{sec:methods}

In this section, we provide a brief review of the adopted neural network architecture and highlight any changes of the networks with respect to the original designs \cite{li2020fourier, lorsung2024physics}.  
FNO, and by extension PITT FNO, aim to learn the operator $\mathcal{G}_\theta : \mathcal{A} \rightarrow \mathcal{U}$, where $\theta$ are the model parameters, $\mathcal{A}$ is the input function space, usually the initial state of the gas flow, and $\mathcal{U}$ is the solution function space, the gas flow evolved to time $t$. Here, these networks are adopted for interpolation. Therefore, the models $\mathcal{M}$ are trained for the task as $I_t = \mathcal{M}(I_0,I_1,t)$, where $I_0$, $I_1$ are the input frames, $t\in[0,1]$ denotes the interpolation time between the input frames and $I_t$ is the interpolated frame.

\subsection{Fourier neural operator}
Neural operators are a type of architecture designed to learn the mapping between infinite-dimensional function spaces, as opposed to finite-dimensional Euclidean spaces. As a result, the neural operator output is discretization invariant, which means that it is independent of the input grid resolution and can be evaluated at any point in the output domain. FNO is a neural operator algorithm that achieves this by processing the data in the Fourier domain, where the frequency components are scale-agnostic. The function mapping of neural operators makes them able to generalise to an entire family of PDEs, as opposed to numerical solvers which solve only on instance of the equation. FNO is successful in modelling turbulent flows compared to other learning-based solvers and provides a significant reduction in computation time compared to traditional PDE solvers \cite{li2020fourier}.

\subsubsection{Fourier layer}\label{subsec:fourier}
The first step in the network lifts the input $a \in \mathcal{A}$ to a higher-dimensional space through a linear transformation layer $P$. Afterwards, the resulting output $v_0$ is passed through several update layers. The updates $v_t \rightarrow v_{t+1}$ are defined as
\begin{equation}
    v_{t+1}(x):= \sigma\big(Wv_t(x)+(\mathcal{K}(a;\phi)v_t)(x)\big)
    \label{eq:update}
\end{equation}
where $W$ is a linear transformation, $\sigma$ is a local non-linear activation function, and $\mathcal{K}$ is a non-local integral operator. The output of these update layers $v_T$ is transformed to the target dimension by local transformation $Q$.

It is the integral operator $\mathcal{K}$ that allows the transfer of information throughout the domain. The integral operator $\mathcal{K}$ is chosen as a kernel integral transformation, parameterised by a neural network. The kernel integral operator mapping is defined by
\begin{equation}
    \big(\mathcal{K}(a;\phi)v_t\big)(x) := \int_D \kappa(x,y,a(x),a(y);\phi)v_t(y)dy
    \label{eq:kernel}
\end{equation}
where $\kappa_\phi$ is the neural network parametrised by $\phi$, which are trainable model parameters. The kernel integral operator can be replaced by a convolution operator in Fourier space through the Fourier transform $\mathcal{F}$. The Fourier integral operator can be redefined as
\begin{equation}
    \big(\mathcal{K}(\phi)v_t\big)(x)=\mathcal{F}^{-1}\big(R_\phi\cdot(\mathcal{F}v_t)\big)(x)
\end{equation}
where $R_\phi$ is the Fourier transform of a periodic function $\kappa$, which is parametrised as a weight tensor of Fourier modes $k$, with the Fourier modes truncated at $k_{\rm max}$. Considering that the domain is a discrete grid of points, the multiplication by the weight tensor $R$ becomes,
\begin{equation}
    \big(R\cdot(\mathcal{F}v_t)\big)_{k,l} = \sum^{d_v}_{j=1}R_{k,l,j}\big(\mathcal{F}v_t\big)_{k,j}, \qquad k=1,...,k_{\rm max}, \quad j=1,...,d_v
\end{equation}
When the discretization is uniform, the Fourier transform $\mathcal{F}$ can be replaced by the Fast Fourier Transform. Most of the computing cost lies in computing the Fourier transform and its inverse; therefore, using the Fast Fourier transform allows for an efficient computation of equation \ref{eq:kernel}. By learning the mapping between Fourier modes, the solution is resolution-invariant.

\subsection{Physics informed token transformer}
While FNO shows impressive results on a number of benchmarks, its purely data-driven and lacks any information on the physics at play. PITT is a transformer model in which tokenized PDEs are embedded in the learning process \cite{lorsung2024physics}. In this way, knowledge about the underlying physical process is fused into neural operator learning, simplifying the learning process. To include information on the underlying PDE the equation is split into its symbols and tokenized. The input of the network consists of the tokenized equations together with the numerical values, grid spacing, and the explicit time difference between simulation steps. The encoded time difference allows output to arbitrary timesteps. The numerical values and the grid are passed through a neural operator (such as FNO, DeepONet or OFormer), which outputs some embedded prediction. This output is passed to a linear attention update module to construct an update operator $F_P$ to the neural operator output.

\subsubsection{Token embedding}
Embedding tokenized equations simplifies the learning process of the model operator $\mathcal{G}_\theta$. The tokenization process occurs by assigning a token to the split symbols, which includes decimal points, commas, and boundary conditions \cite{charton2020learning}. The tokens are arranged in a list in the order of the equation symbols, such as 
\begin{equation}
    \begin{split}
        \frac{\partial}{\partial t}e(x, y, t)&=\text{Derivative}(e(x, y, t),t)\\ &= [\text{Derivative},(,e,(,x,,,y,,,t,),,,t,)] \\ &=[8,0,36,0,14,37,16,37,11,1,37,11,1],
    \end{split}
\end{equation}

with the target time value also being appended at the end. The tokenized equations are used to construct the key, query, and value matrices in a Multi Head Attention block with a set of learnable weight matrices. The Multi Head Attention block uses self-attention to create an enhanced output embedding of the PDE \cite{vaswani2017attention}. This process can be written as
\begin{equation}
    \text{Attention}(Q, K, V) = \text{softmax}\bigg(\frac{QK^{\intercal}}{\sqrt{d_k}}\bigg)V
\end{equation}
where $Q$, $K$, and $V$ are the query, key, and value matrices, and $d_k$ the length of the query and key. PITT learns the relation between the symbols of the equation during the training process, and these relations are captured in the latent equation representation that is output by the self-attention block. This latent representation is used to construct the key and query for the following linear attention block.

\subsubsection{Linear attention update}

The function of the linear attention update module is to add information of the underlying PDE to the neural operator output. The module uses this, together with time embedding, to construct an update function to the neural operator output. 
The output of the token transformer module is used to construct the keys and queries for the module. The value matrix is provided by embedding the neural operator output.

The linear attention update module consists of a linear attention block and a numerical update block. Unlike standard self-attention, linear attention does not use the softmax function \cite{cao2021choose} and instead approximates standard attention through a feature map, achieving better scalability with the sequence length.

Through linear attention, the neural operator output is infused with the embedding of the PDE. The output of this block is stacked with the time embedding and is passed through a fully connected projection layer, after which it is added to the value matrix. This process can be denominated as $V_l \leftarrow V_{l-1} + \text{MLP}([X_l,t_l])$, where $V$ is the value matrix, $X$ is the linear attention output, $t$ the time embedding, and $l$ denotes the layer index.

The module output is obtained by unembedding the value matrix of the last layer back to the target dimensions, which is added to the neural operator output to create the target prediction.

\subsubsection{Summary of PITT adaptation}
\begin{figure}
    \centering
    \includegraphics[width=\linewidth]{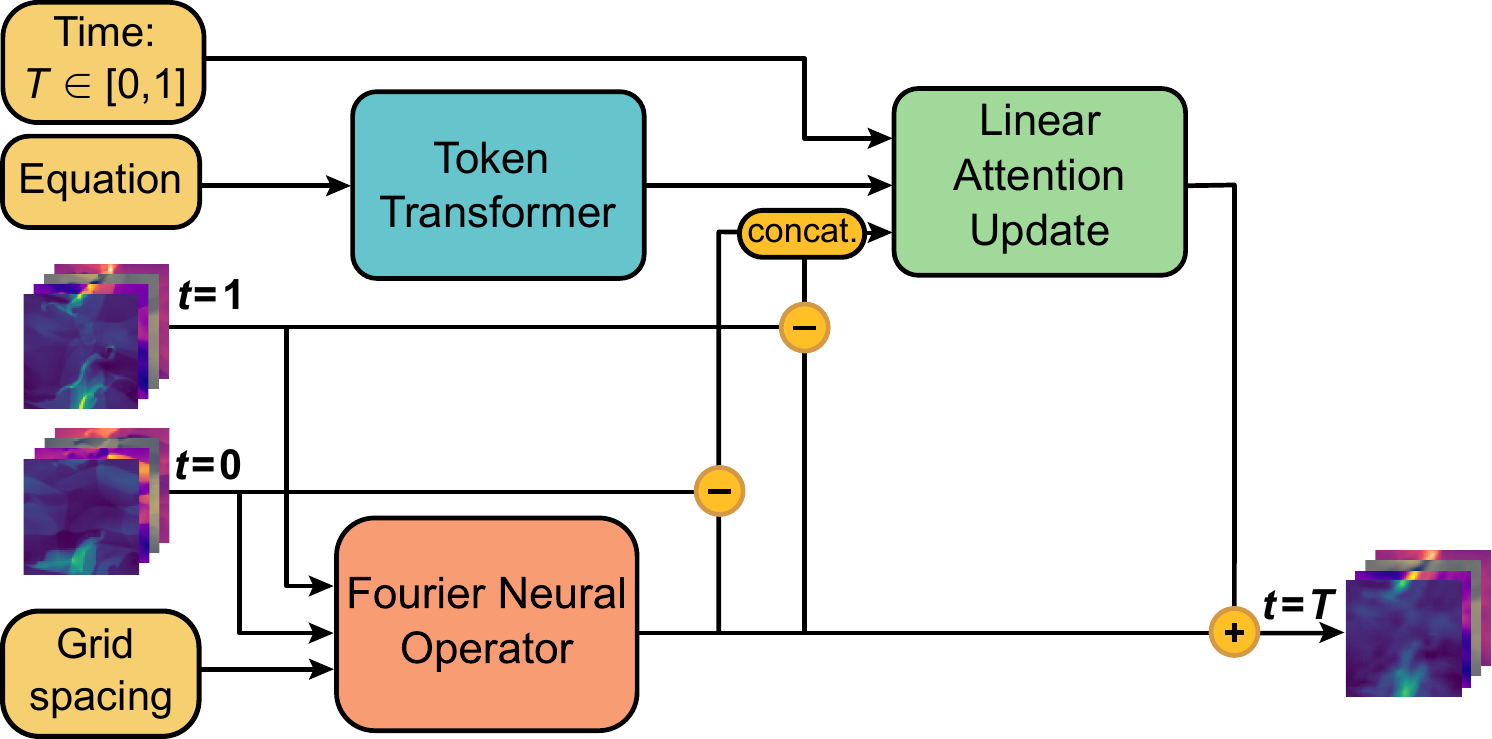}
    \caption{The adapted PITT network uses the tokenised interpolation time and equation to create an update to the neural operator output. The start ($t=0$) and end ($t=1$) frame are both independently subtracted from the neural operator output and concatenated before input in the attention block to create symmetry in the network predictions. The attention block explores cross-correlations in the multi-channel data, which is used together with the equation and time embedding to update the neural operator output to target time $T$.}
    \label{fig:adapted_PITT}
\end{figure}

In order to use PITT for the interpolation task, a number of adaptations were made to the original design. Additionally, the network is changed to exploit cross-channel correlations in multi-channel data. The adaptations are summarised here.
\begin{itemize}
    \item The time embedding is changed to the interpolation time ($t\in[0,1]$, where $t=0$ and $t=1$ denote the start and end frame times, respectively) as opposed to the time difference between steps
    \item Both the start as the end frame are subtracted from the FNO output and concatenated before the attention block to create symmetry in the network prediction
    \item The linear attention update embedding layer is extended to allow for multi-channel data to construct the value matrix from the neural operator output
\end{itemize}
In addition to these design changes, several changes were made in data preparation and selection. The adapted PITT architecture is shown in Figure \ref{fig:adapted_PITT}.

\section{Training}\label{sec:training}
\subsection{Setup}
Model training is conducted using PyTorch. During the PITT training process, we use Adam optimizer with a learning rate of $10^{-4}$ for 50 epochs with a batch size of 16. FNO uses the same setup that runs for 35 epochs, with the loss curve being observed to flatten at the end of the training process. The networks are trained using an NVIDIA RTX6000 Ada GPU, with FNO taking approximately 3 hours and PITT FNO taking approximately 5 hours to train. The 2D dataset governed by the Euler equations is divided between training, validation and test using 70\%, 18\% and 12\% of the available data, respectively.

\subsection{Data}
For complex multi-channel fluid data in two dimension, data was selected from the public databank \textit{The Well}, which provides numerical simulations for the purpose of training ML networks \cite{mandli2016clawpack}. The selected dataset is that of the Euler equations for a compressible gas,
\begin{equation}
    U_t + F(U)_x + G(U)_y = 0
    \end{equation}
where
\begin{equation}
    \begin{aligned}
    &U = \begin{bmatrix}
        \rho \\
        \rho u \\
        \rho v \\
        e
    \end{bmatrix}, \quad
    F(U) = \begin{bmatrix}
        \rho u \\
        \rho u^2 + p \\
        \rho u v \\
        u(e + p)
    \end{bmatrix}, \quad
    G(U) = \begin{bmatrix}
        \rho v \\
        \rho u v \\
        \rho v^2 + p \\
        v(e + p)
    \end{bmatrix}, \\
    &e = \frac{p}{\gamma - 1} + \rho \frac{u^2 + v^2}{2}
\end{aligned}
\end{equation}
where $\rho$ is the density, $u$ and $v$ are the velocity components in the $x$ and $y$ directions, $e$ the energy, $p$ the pressure, and $\gamma$ the adiabatic index. The data contains 4 channels: density, energy, and the $x$ and $y$ components of the momentum. The gas constant has values of $\gamma\in\{$1.13, 1.22, 1.3, 1.33, 1.365, 1.4, 1.404, 1.453, 1.597, 1.76$\}$. For every value of $\gamma$, the dataset contains 500 simulations, for a total of 10000 simulations that run for 100 timesteps. The dimensions of the discretized data are $512\times 512$, and are downsampled to $64\times 64$ for training and testing. Simulations can contain turbulence, moving shocks and contact discontinuities, making the dynamics especially challenging to learn. Periodic boundary conditions are used which allows us to monitor strict conservation.  

\section{Results}\label{sec:results}

In this section we present the performance of PITT FNO on data governed by the Euler equations. The quantified performance is compared to a stand-alone FNO benchmark and to linear interpolation for a range of input frame separations. Furthermore, we will investigate the ability of the network to evolve simulations continuous over time. To conclude, we will perform a spectral analysis to probe the performance at small spatial scales. 

\begin{figure*}[htbp]
    \centering
    \includegraphics[width=\linewidth]{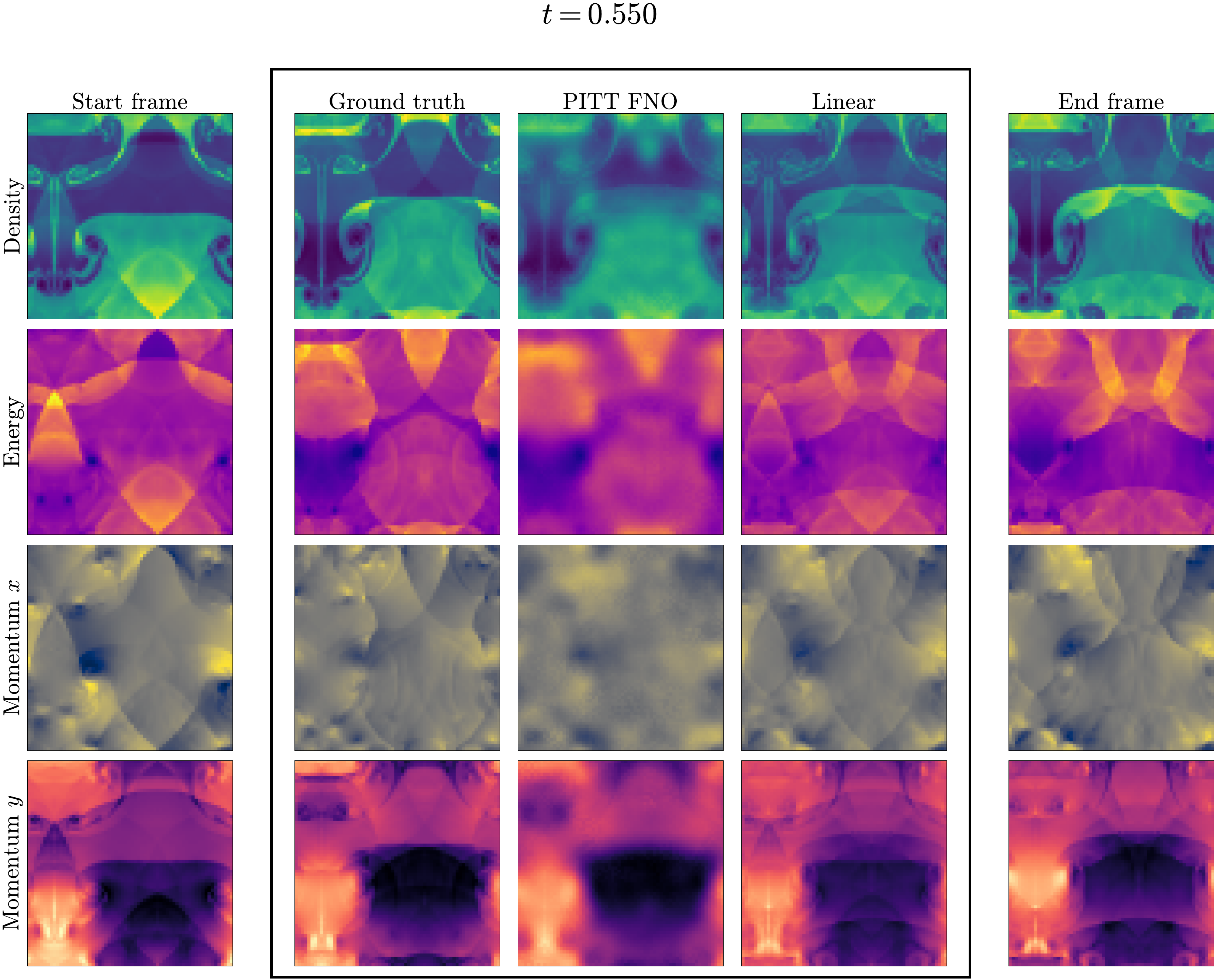}
    \caption{Example of PITT FNO prediction for the Euler dataset. \textbf{From top to bottom:} Density, energy, momentum in $x$, momentum in $y$. The start frame ($t=0$) and end frame ($t=1$) are input frames into the network and are separated by 20 data frames, with the target being at interpolation time $t=0.55$. The PITT FNO prediction is compared to the target frame and linear interpolation at the interpolation time. By learning the dynamics between snapshots through the PDE embedding, PITT FNO is able to predict non-trivial features such as crossing shock fronts observed in the energy prediction. For a movie equivalent of this Figure, go to this \href{https://youtu.be/JbK104xy74Y?si=sANgVU_EAeiiMg3l}{link}.}
    \label{fig:PITT2D-test}
\end{figure*}

\subsection{Comparison to baseline}
The PITT FNO model is extended to support multi-channel data and is therefore able to perform simultaneous predictions across all physical domains, while allowing the network to exploit cross-channel correlations. Figure \ref{fig:PITT2D-test} demonstrates that PITT FNO is capable of evolving coherent structures on a wide range of spatial scales, even at large temporal separations. 
It can be appreciated how linear interpolation just 'blends' start- and end-frame, resulting in sharp but incorrect predictions.  By contrast, PITT FNO learns the dynamics between the snapshots which leads to non-trivial predictions.  For example, PITT FNO is able to capture the crossing shock fronts and recovers the wedge-shaped high-energy feature near the upper boundary in the target frame.   

\begin{figure}[h]
    \centering
    \includegraphics[width=\linewidth]{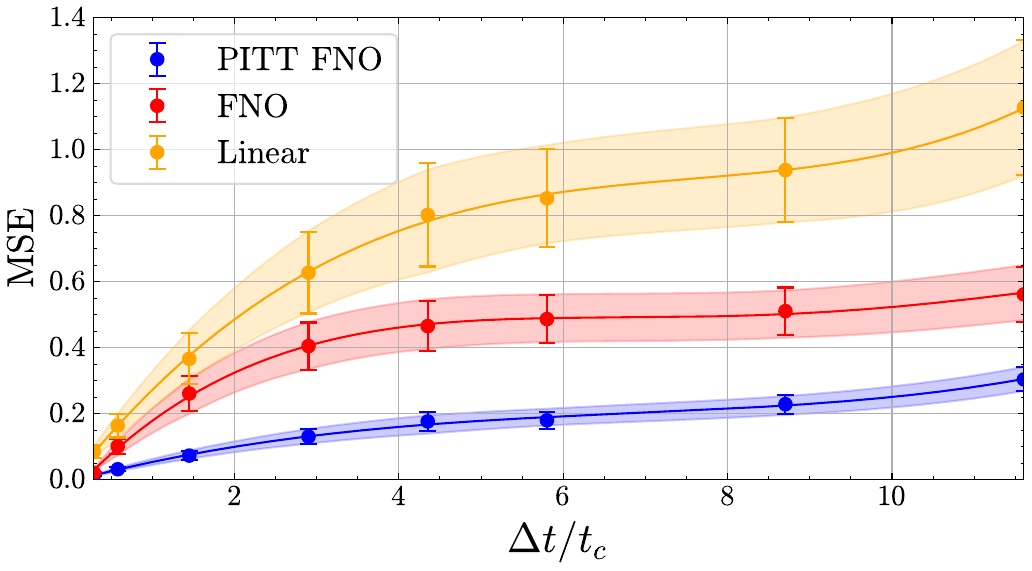}
    \caption{Mean square error as a function of separation between the input frames for the Euler equations. PITT FNO is able to outperform stand-alone FNO and linear interpolation across all levels of separation. Compression ratios between PITT FNO and linear interpolation range from 6 for low MSE (0.1) to 10 at higher MSE (0.2-0.3). All separations and compression ratios are given in Table \ref{tab:MSE-Euler}.}
    \label{fig:PITT2D-euler}
\end{figure}

For a quantitative comparison, the performance of PITT FNO and linear interpolation are expressed as the mean square error (MSE). Training is repeated across a range of input frame separations and evaluated on the test set, the results of which are presented in Figure \ref{fig:PITT2D-euler}. Here, input frame separation is expressed in correlation time ($t_c$), which is explained in more depth in Section \ref{sec:correlation-time}. 

PITT FNO consistently outperforms the baseline in all separation levels. More importantly, for every chosen MSE value, we achieve a sizeable increase in the separation between the input frames with PITT FNO compared to linear interpolation. The separation achieved by the models for selected error values is shown in Table \ref{tab:MSE-Euler}. The compression ratio, the ratio between the separations achieved by PITT FNO and linear interpolation for the same error value, is shown in the right-most column. Compression ratios between PITT FNO and linear interpolation range from $\sim$6-10, indicating that PITT FNO can achieve comparable performances with significantly fewer frames. At small MSE, the corresponding compression ratios are lowest. As linear interpolation performance drops off significantly at higher separations, compression ratios increase sharply for higher MSE.

\begin{table*}[h]
\centering
\caption{Maximum separation in number of frames to achieve a desired MSE for the Euler equation dataset for the different methods evaluated. Right-most column shows the compression ratio between PITT FNO and linear interpolation at that error value.}
\label{tab:MSE-Euler}
\begin{tabular}{|c||c|c|c|c|}
\hline
\multirow{2}{*}{\textbf{MSE}} & \multicolumn{3}{c|}{\textbf{Separation}} & \textbf{Compression ratio}\\ \cline{2-5}
 & \textbf{Linear} & \textbf{FNO} & \textbf{PITT FNO} & \textbf{PITT FNO$\big/$Linear}\\ \hline
0.1 & 2.37 & 3.98 & 14.74 & 6.22 \\ \hline
0.2 & 5.07 & 7.71 & 48.30 & 9.53 \\ \hline
0.3 & 8.04 & 12.72 & 79.01 & 9.83 \\ \hline
0.4 & 11.30 & 19.68 & 89.26 & 7.90 \\ \hline
0.5 & 15.13 & 51.25 & 98.87 & 7.61 \\ \hline
\end{tabular}

\end{table*}

\subsection{Time evolution}
\begin{figure*}[htbp]
\centering
    \includegraphics[width=\linewidth]{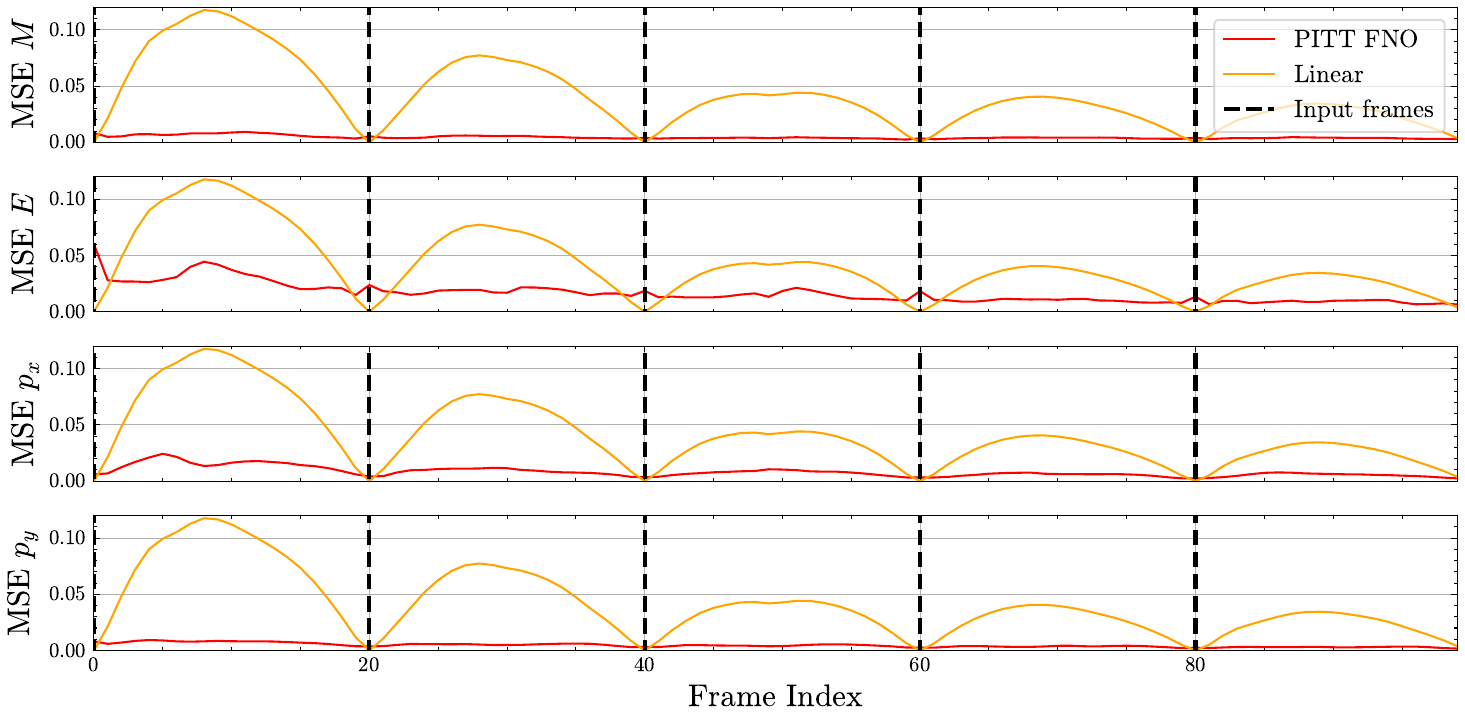}
    \caption{Mean square error split up between variables during a single simulation. \textbf{From top to bottom:} Density, energy, momentum in $x$, momentum in $y$. The dashed black line shows the location of the input frames used to predict the fluid state between the inputs, with separations of 20 frames. Compared to linear interpolation, PITT FNO provides remarkably consistent predictive power across the entire domain.}
    \label{fig:loss}
\end{figure*}

\begin{figure*}[htbp]
    \centering
    \includegraphics[width=\linewidth]{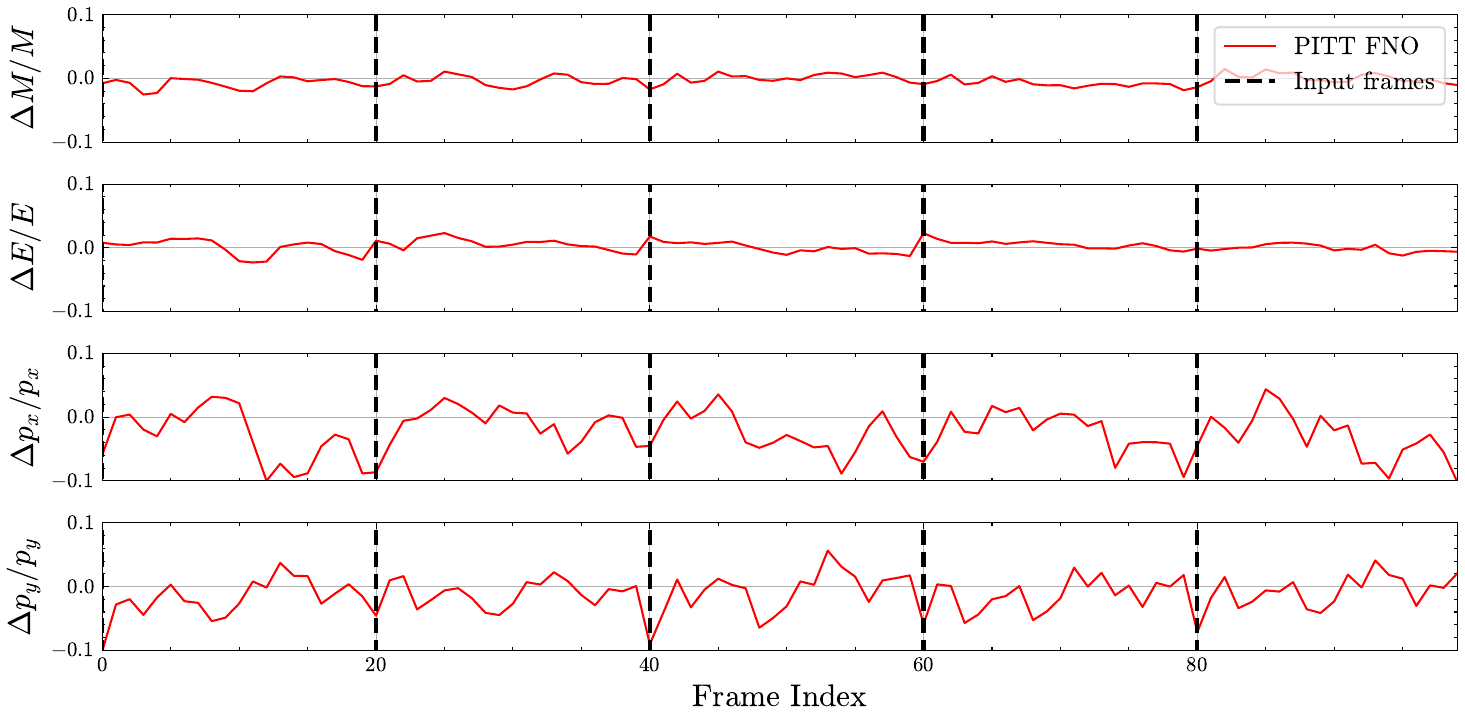}
    \caption{Relative error in mass, energy and momentum conservation during a single simulation. \textbf{From top to bottom:} Density, energy, momentum in $x$, momentum in $y$. The dashed black line shows the location of the input frames used to predict the fluid state between the inputs, with separations of 20 frames. Likely as a result of PDE embedding, errors in mass and energy conservation are low throughout the simulation.}
    \label{fig:cons-values}
\end{figure*}

The time-continuous nature of PITT allows for an analysis of how errors evolve over time during a single simulation. Figure \ref{fig:loss} shows this error progression in the case of 20-frame input frame separation. As expected in a network with symmetry around the input frames, the errors decrease closer to the input frames. Although linear interpolation performs well in the vicinity of the input frames, PITT FNO consistently outperforms linear interpolation across the entire domain. Beyond standard error metrics, we evaluated the physicality of the interpolated outputs through the conservation of mass, energy, and momentum. Significant deviations of these quantities can undermine the reliability of the simulations. Taking the same simulation as previously shown, Figure \ref{fig:cons-values} reports the relative errors of conserved quantities during the simulation. The model demonstrates excellent conservation of mass and energy, with errors below 3\% throughout the simulation. Both momentum channels show more variability, with errors up to 10\% early in the simulation. The relatively small errors in mass and energy conservation could indicate that the network benefits from the physics-informed contribution of PITT.

\subsection{Spectral analysis}
\begin{figure}[ht]
    \centering
    \includegraphics[width=\linewidth]{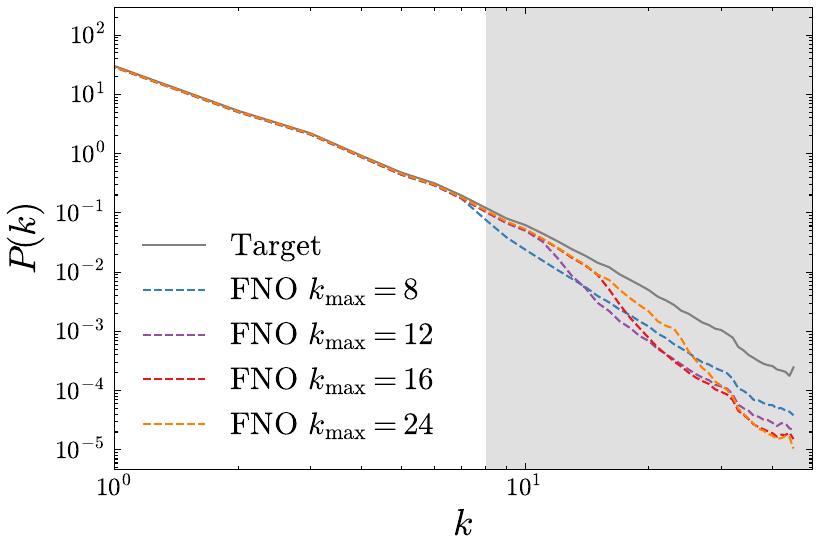}
    \caption{Power spectrum of the velocity field and the predicted spectrum of various FNO iterations. FNO iterations with a higher spectral cut-off maintain better performance at higher wavenumbers, although there is a general deviation from the target at higher wavenumbers.}
    \label{fig:power_spectrum}
\end{figure}

Although the adapted PITT FNO reproduces the large-scale dynamics of the Euler equations, we observe that its predictive power drops off at finer scales. To investigate how network parameters affect fine scale performance, we perform a spectral analysis of the model’s predictions. As described in Section \ref{subsec:fourier}, FNO filters out a set of high-frequency modes during the linear transformation to preserve the compactness of the model. We therefore hypothesise that retaining more modes in the Fourier layers might restore resolution at smaller spatial scales. Allowing FNO to learn the dynamics of these high-frequency modes in the Fourier domain could improve predictions at these scales. To test this, we trained and evaluated multiple iterations of FNO with progressively larger spectral cut-offs. The power spectrum shows how energy is distributed across spatial scales and is given by
\begin{equation}
P(k)=\sum_{|\textbf{k}|=k}|\hat{f}(\textbf{k})|^2
\end{equation}
where $\hat{f}(\textbf{k})$ is the Fourier transform of a variable and $k=\sqrt{k_x^2+k_y^2}$ is the wavenumber. Figure \ref{fig:power_spectrum} shows how the different FNO iterations compare to the ground truth for the power spectrum of the velocity field at low separation. Here, FNO reproduces the target spectrum well at low wavenumbers, corresponding to large-scale structures. At high wavenumbers, the FNO spectra start to drop below the target spectrum, indicating a decrease in predictive performance. FNO experiments with higher spectral cut-offs mitigate this divergence, suggesting that extending the retained frequency range does improve the model’s ability to capture small-scale dynamics. This decoupling from the target spectrum occurs at $k\approx10$. 
Beyond $k=10$, each $k_{\rm max}$ experiment matches those performed with higher $k_{\rm max}$ until it breaks away at the mode truncation ($k_{\rm max}$). However, compared to the ground-truth, even the  $k_{\rm max}=24$ experiment misses structure well below the truncation (around $k>10$), indicating that the mode truncation is not the only reason for the loss of fine detail. 

\subsection{Correlation time}\label{sec:correlation-time}
\begin{figure}
    \centering
    \includegraphics[width=\linewidth]{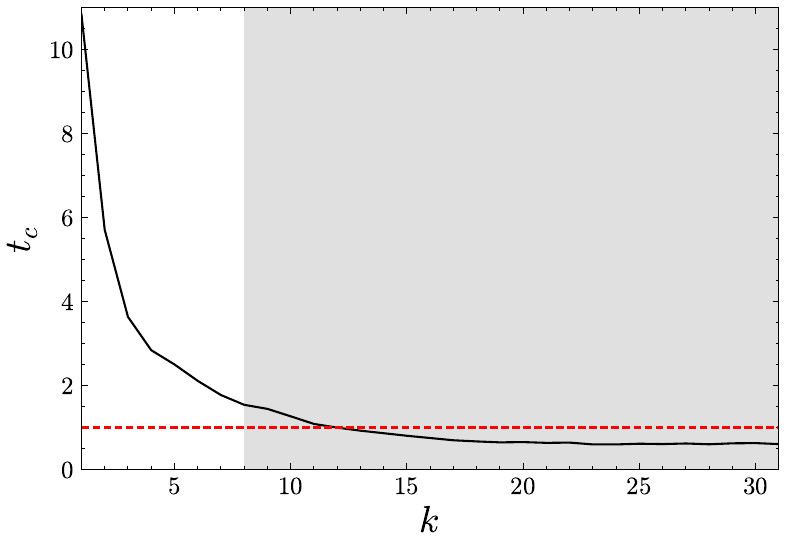}
    \caption{The correlation time as a function of the wavenumber $k$. The correlation time quickly decreases for smaller spatial scales, falling below the temporal resolution of the data (dashed red line) around $k=12$. Dynamics finer than this scale are therefore poorly sampled, limiting predictive strength.}
    \label{fig:correlation-time}
\end{figure}

Although extending the FNO network to higher wavenumbers can assist with the performance at fine scales, it does not fully recover the structure of the ground truth. To further analyse this issue, in this section we investigate how the performance at a given scale depends on the \textit{temporal sampling} of the data. 
To provide context for the timescales, we first compute the scale-dependent correlation time of the input data. The correlation time is defined as the point where the correlation coefficient is equal to $1/e$. The correlation coefficient $\mathcal{C}$ is a quantitive measurement of the similarity between two fields $f(x,y)$ and $g(x,y)$, and is calculated as 
\begin{equation}
\mathcal{C} = \frac{1}{N\sigma_f\sigma_g}\sum_{x,y}([f(x,y)-\mu_f][g(x,y)-\mu_g])
\end{equation}
where $\mu_f$, $\mu_g$ correspond to the means of $f$ and $g$, $\sigma_f$, $\sigma_g$ to the standard deviations, and $N$ to the number of pixels. When the input frame separation exceeds the correlation time at the spatial scale, the dynamics becomes poorly sampled. The correlation time drops significantly at smaller spatial scales shown in Figure \ref{fig:correlation-time}, resulting in the cadence of the training data dropping below the correlation time at $k\approx12$. 

Note that the convergence of the correlation time for higher wavenumbers is an artefact of the coarse temporal resolution of the data, and the value of $\sim 0.6$ is the minimal correlation time that can be calculated.
\begin{figure}[h]
    \centering
    \includegraphics[width=\linewidth]{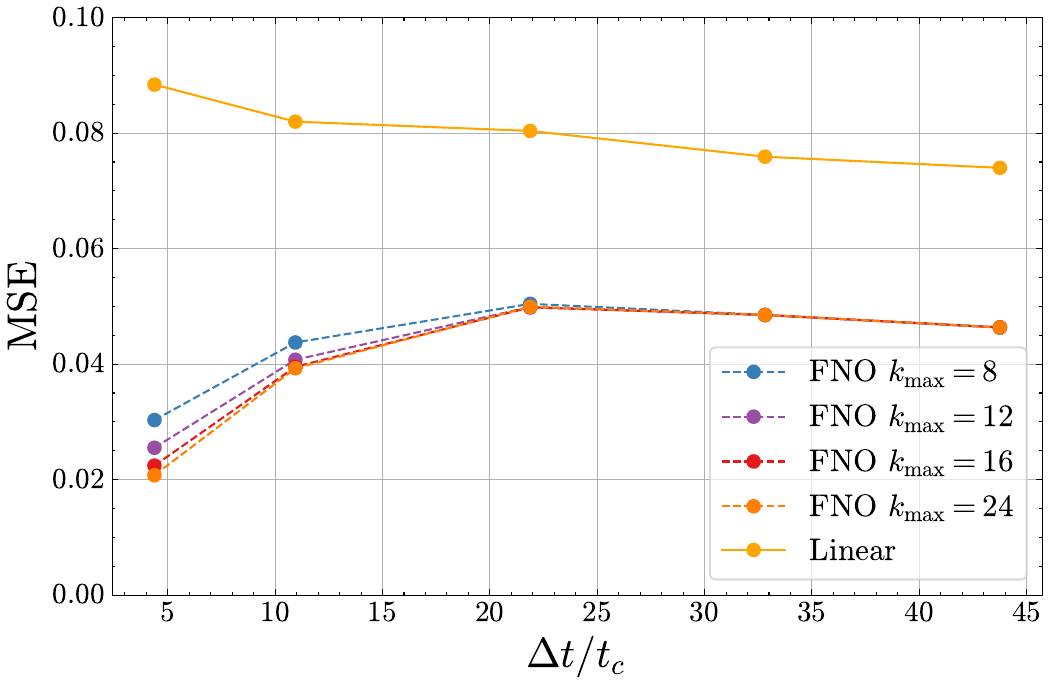}
    \caption{Performance of FNO models with varying spectral cut-offs in the high-frequency band, as a function of input frame separation. Iterations with higher spectral cut-offs outperform other iterations at low separations where input frames are correlated, but this difference disappears at higher separations where input correlations drop to zero.}
    \label{fig:fno-modes}
\end{figure}

The effect of frame separation exceeding the correlation time can be seen when we focus exclusively on the high-frequency band shown in grey in Figures \ref{fig:power_spectrum} and \ref{fig:correlation-time}. Figure \ref{fig:fno-modes} shows the MSE of the FNO iterations in that high-frequency band as a function of the separation between the input frames. The Figure demonstrates that for smaller levels of separation between inputs, adding more frequency modes to the FNO learning improves performance. However, for higher levels of separation this improvement disappears, with the performance plateauing at MSE$\approx$0.045-0.05. 
Once the input frames no longer sample the dynamics at small scales scales, the network assigns near-zero coefficients to these modes, gaining no advantage from added capacity. FNO therefore completely relies on PITT to learn the dynamics at finer scales through the PDE embedding.

The plateauing around 10-20 correlation times is likely a result of the undersampling of small scale dynamics as the network is essentially fed with uncorrelated noise. It is encouraging though that this plateauing happens ``late'' and FNO learns small scale dynamics up to a separation on the order of $10$ correlation times.  

A similar limitation appears to constrain the combined PITT FNO network. As PITT FNO has time-continuous capabilities, it should be able to learn the dynamics at smaller scales better than stand-alone FNO. However, as shown in Figure \ref{fig:PITT2D-test} also PITT FNO is lacking features at small scales and might suffer from the temporal undersampling issue.

To summarize the investigation of the spatial and temporal spectra: PITT FNO learns large-scale motion well while struggling with small-scale structure. The results of this section suggest that the spatial resolution of the model output is constrained by the temporal resolution of the training set.

\section{Discussion}\label{sec:discussion}
We show that PITT FNO requires $\sim$6-10 times less data than linear interpolation to achieve the same MSE, tested on a dataset governed by the Euler equations. While MSE provides a convenient quantitative measure of reconstruction accuracy, it does not guarantee physical consistency. In particular, low MSE values may coincide with violations of conservation laws or smoothed-out discontinuities. We therefore complement MSE with diagnostics of mass, momentum, and energy conservation, as well as spectral analyses to assess the physical fidelity of the interpolated fields. 
To estimate the impact on a particular application, we consider the slow-light raytracing simulations of \cite{vos2024magnetic}. Their implementation required $\sim$600 snapshots ($\sim$350 GB RAM), using nearest-neighbour interpolation due to the high time resolution. Assuming a PITT FNO compression ratio of 10, storage needs would drop tenfold, reducing memory usage to $\sim$35 GB, thus freeing up space for longer or more complex simulations. Alternatively, instead of improving compression, PITT FNO could be used to obtain more accurate predictions while retaining the number of snapshots. When using the trained PITT FNO model for inference, generating a single frame in the Euler dataset takes approximately $t=0.005\;\text{s}$. As the dataset has a resolution of 64$\times$64 with 4 variable channels, this is equivalent to an inference time of approximately $t=3\times 10^{-7}\;\text{s/pixel/variable}$.  In a typical raytracing application \citep[e.g.][]{PratherDexterEtAl2023}, around 100 interpolation steps are performed per geodesic at a resolution of e.g. $400^2$ rays and 8 interpolated variables.  This would result in an interpolation time of $\approx 40\rm s$ which is on the order of the runtime of current ``fast-light'' models of $\sim 30\rm s$.  Hence adding ML interpolation capabilities at runtime is already feasible at comparable cost.  

Despite promising results, several challenges remain. Our evaluation reveals a decline in performance within the high-frequency range. Similar limitations of FNO at fine scales have been reported \cite{duarte2025spectral}, and we show that it cannot be fully resolved by extending the truncation of Fourier modes in FNO. In addition, increasing the temporal separation between the input frames further degrades performance at fine scales and mitigates any gains that could be made by extending mode truncation. The PITT component can alleviate some of these problems through its continuous in time learning capabilities. However, the combined PITT FNO network remains constrained by the temporal resolution of the training data, which for the data used exceeds the correlation time for wavenumbers above $k=12$. Several studies have emphasized the need for high-resolution training data to accurately predict fine-scale features \cite{kim2021unsupervised, fukami2023super}, suggesting a direct correlation between model performance and training data resolution. To probe the impact of training resolution on fine-scale performance, more analyses are required. A number of models have introduced various strategies to improve predictive power in the high-frequency regime \cite{liu2020deep, wang2020towards, fukami2021machine}. However, achieving consistent improvements at these scales remains a significant challenge. A better understanding of how temporal resolution constrains model performance could give valuable insight to improve fine-scale model accuracy.

\section{Conclusion}\label{sec:conclusion}
In this work, we present our findings on the application of existing machine learning architectures for temporal interpolation. This is done with the objective of increasing the cadence of fluid simulations in a post-processing step. Interpolation schemes can be useful whenever the cadence required exceeds the output frequency or when simulations are restricted by memory or storage space. The model evaluated in this paper is a modified version of PITT FNO, a network originally designed for extrapolating fluid simulations. We adapt PITT FNO for multi-channel interpolation tasks and demonstrate that it offers several desirable properties:
\begin{enumerate}
    \item An effective interpolation separation that is approximately 6-10 times higher than that achieved by linear interpolation for the same performance
    \item Strong adherence to mass and energy conservation laws
    \item Multi-channel support and continuous-in-time interpolation capabilities
\end{enumerate}
These findings show the potential of using physics-informed networks for interpolation applications. However, there are still limitations that need to be addressed. Most notably, the degradation of model performance at smaller scales.

\section*{Acknowledgements}
Simulations have been carried out using the HIPSTER cluster of the University of Amsterdam. We thank Daniela Huppenkothen and Jesse Vos for insightful discussions. JD is supported by NASA through the NASA Hubble Fellowship grant HST-HF2-51552.001A, awarded by the Space Telescope
Science Institute, which is operated by the Association of Universities for Research in Astronomy, Incorporated, under NASA contract NAS5-26555



\bibliographystyle{elsarticle-num} 
\bibliography{refs}


\end{document}